\documentclass[runningheads]{llncs}
\usepackage{graphicx}
\usepackage{booktabs}
\usepackage[ruled,noresetcount,noend]{algorithm2e}
\usepackage{algpseudocode}
\usepackage{booktabs} % For formal tables
\usepackage{mathtools}
\usepackage{caption}
\usepackage{subcaption}
\usepackage[sort&compress,numbers]{natbib}
\newtheorem{condition}{C}
\begin{document}
\title{A two-level solution to fight against dishonest opinions in recommendation-based trust systems}
\author{Omar Abdel Wahab\inst{1} \and
Jamal Bentahar\inst{2} \and
Robin Cohen\inst{3} \and
Hadi Otrok\inst{4} \and
Azzam Mourad\inst{5}}
\authorrunning{Omar Abdel Wahab et al.}
\institute{Department of Computer Science and Engineering, Universit\'{e} du Qu\'{e}bec en Outaouais \\ \email{omar.abdulwahab@uqo.ca} \and
Concordia Institute for Information Systems Engineering, Concordia University \\ \email{bentahar@ciise.concordia.ca} \and David R. Cheriton School of Computer Science, University of Waterloo \\ \email{rcohen@uwaterloo.ca} \and Department of ECE, Khalifa University of Science and Technology \\ \email{Hadi.Otrok@ku.ac.ae} \and Department of Computer Science and Mathematics, Lebanese American University \email{azzam.mourad@lau.edu.lb}}
\maketitle              % typeset the header of the contribution
\begin{abstract}
In this paper, we propose a mechanism to deal with dishonest opinions in recommendation-based trust models, at both the collection and processing levels. We consider a scenario in which an agent requests recommendations from multiple parties to build trust toward another agent. At the collection level, we propose to allow agents to self-assess the accuracy of their recommendations and autonomously decide on whether they would participate in the recommendation process or not. At the processing level, we propose a recommendations aggregation technique that is resilient to collusion attacks, followed by a credibility update mechanism for the participating agents. The originality of our work stems from its consideration of dishonest opinions at both the collection and processing levels, which allows for better and more persistent protection against dishonest recommenders. Experiments conducted on the Epinions dataset show that our solution yields better performance in protecting the recommendation process against Sybil attacks, in comparison with a competing model that derives the optimal network of advisors based on the agents' trust values.
\end{abstract}
\section{Introduction}
\label{Sec:Intro}
Collecting recommendations is an important step of any trust establishment process. It involves consulting a set of agents (often referred to as advisors) regarding the behavior of other agents. Therefore, ensuring the authenticity of the collected recommendations is essential to the success of trust modeling. Thus, it is of prime importance to reason about whose opinions to elicit when collecting recommendations and under what circumstances \cite{sen2013comprehensive}. %In other words, the recommendation engine should be able to wisely choose the agents with which to consult so as to obtain credible recommendations.
This can be done either at the collection level by pre-filtering the set of agents to whom recommendation requests should be sent or at the processing level by deciding (out of the collected recommendations) whose opinions to keep and whose opinions to discard or discount. The goal of this work is to come up with a reasoning model that helps trust modelers ensure the authenticity of recommendations at both the collection and processing levels. We argue first that pre-filtering agents at the collection level is quite hard especially in large-scale dynamic settings. In fact, with the growing number of deployed agents (e.g., smart vehicles, robots, etc.), it is becoming quite hard to design a filtering algorithm that can efficiently cover all the existing agents. In addition, the behavior of the agents is often subject to change (i.e., being honest sometimes and dishonest some other times) especially in the presence of smart malicious agents that seek to avoid being detected and punished. Therefore, we chose, in the first part of our solution, to allow agents to self-decide on whether to participate in the recommendation process or not. The decision of the agents is mainly affected by two factors. The first is the amount of resources they can allocate towards trust modeling. For instance, in some applications such as intelligent transportation systems, autonomous cars might prefer to fully dedicate their resources to making real-time driving decisions. The second factor is the accuracy level that agents believe their recommendations have. At this stage, we argue that only honest agents are expected to follow this self-withdrawal principle to both help the decision-making process and maintain their own credibility. On the other hand, dishonest agents are often encouraged to still participate by giving untruthful opinions to manipulate the recommendation decisions. Therefore, further reasoning is needed at the precessing/aggregation level to discard such dishonest opinions. The challenge here is to design an aggregation solution that is resilient to dishonest agents even when such agents form the majority. This is because dishonest agents might take advantage of the self-withdrawal of honest agents to impose their opinions by majority. Worse, dishonest agents might launch some attacks (e.g., Sybil, whitewashing, camouflage, etc.) to further increase their chances of swaying the recommendation decisions.

\subsection{Contributions}
Our solution is composed of four principal phases: recommendation collection, aggregation to discourage dishonesty, credibility update, and participation motivation. In the first phase, agents are asked by the underlying recommender system to run the decision tree machine learning technique on their datasets to derive appropriate recommendations. Specifying the machine learning technique to be used is essential in our solution to guarantee the homogeneity in terms of opinions origins. Leaving this decision for the agents might lead to inconsistent decisions that are biased towards some machine learning algorithms, dataset size, number of dimensions in the dataset, etc. Our choice of decision tree stems from its lightweight nature which makes it suitable for situations wherein resource-constrained agents exist. Interested agents (e.g., those that have enough resources) train the decision tree algorithm on their datasets which record their previous interactions with agents of different specifications and behavior and decide, based on the obtained accuracy level, on whether to submit their opinions or not. Our decisions for this collection phase provide important insights into two of the stages of trust modeling highlighted by Sen \cite{sen2013comprehensive}. The Engage step (who to ask) is moderated by agents having the choice of opting out. The Use step (taking actions based on trust modeling) is also facilitated, as the common reasoning that we require yields insights into the origins of the reputation values provided.

In the second phase, the opinions from different agents are aggregated by the underlying recommender system using the Dempster-Shafer Theory (DST) of evidence \cite{shafer1992dempster} to arrive at final aggregate initial trust scores. The aggregation technique takes into consideration the collusion attacks (i.e., Sybil, camouflage, and whitewashing) that might occur among agents to give dishonest recommendations with the purpose of promoting/demoting some agents and accounts for the special case where dishonest agents are the majority. In the credibility update phase, we propose a mechanism to update the credibility scores of the agents that participate in the recommendation collection phase and ensure the authenticity of the recommendation process \cite{wahab2015survey,wahab2019resource,wahab2020endorsement}. Finally, in the participation motivation phase, we argue that in settings where agents may be both requesters and providers of information (such as service-oriented environments) agents will be particularly motivated to assist when recommendations are requested. We discuss this component of our model in greater detail, later in the paper.

\section{Related Work}
\label{Sec:RelatedWork}
In \cite{guo2014merging}, the authors assume that users have a network of trusted peers and combine the opinions of these advisors, averaging their ratings on commonly rated items. Trust and social similarity are merged to represent the active user's preferences and generate appropriate recommendations, for the case when little is known about the user. In \cite{massa2006trust}, the authors employ collaborative filtering to ease the recommendation process using a two-stage methodology. In the first stage, users are represented  in the form of a social network graph and the task is to collect trust statements regarding newcomer users. In the second stage, all the trust statements are analyzed and aggregated using the averaging technique to predict the trust scores of the newcomer users. In \cite{lika2014facing}, the authors propose a three-phase approach to address the problem of cold-start users. In the first phase, the C$4.5$ and Naive Bayes techniques are employed to assign new users to specific groups. In the second phase, an algorithm is proposed to explore the neighbors of the new user and an equation is presented to compute the similarity between new users and their neighbors in terms of characteristics. In the final phase, a prediction method is used to estimate the final rating of the new user for every existing item, where the rating is a weighted sum of ratings submitted by the user's neighbors on the corresponding item. In \cite{quercia2007trullo}, local information obtained through analyzing users' past ratings is used to assign initial trust values for ubiquitous devices (e.g. smart vehicles). More specifically, this approach uses Singular Value Decomposition, which represents known trust values as linear combinations of numeric features and then combines them to estimate the unknown trust value.

The primary contrast with the aforementioned models can be outlined in three main points. First, the discussed approaches deal with dishonest recommendations at either the collection or processing level. Our solution operates on both levels to increase the protection against dishonest recommendations. Second, in the approaches that operate at the collection level, the choice of evaluating the adequacy of the agents in participating in the recommendation process is left only to the recommender system through computing the optimal network of agents that maximize trust, without accounting for the self-willingness and self-confidence of the agents themselves. In other words, although some agents might be highly trusted in general, this does not mean that they will be providing accurate recommendations for all types of requests. The accuracy here might vary according to the data available to these agents, the characteristics of the agents being recommended, and the technique used to compute recommendations. To tackle this challenge, we leave in this work the choice for the agents to self-asses their own ability in participating in the recommendation process or not. In addition, we force the agents to use a common recommendation computation technique (i.e., decision tree) to increase the homogeneity of the received recommendations. Third, different from the literature which employs aggregation techniques that might be vulnerable to manipulation, we take advantage of DST to perform the aggregation in a manner that is sensitive to possible dishonesty even in extreme cases wherein dishonest agents might be the majority.

\section{Proposed Solution}
\label{Sec:TrustBootstrapping}
\subsection{Solution Overview}
The proposed solution can be summarized as follows. Each agent maintains a dataset which corroborates its previous interactions with agents of different specifications and behavior. This dataset is assumed to be labelled in the sense that the agent would classify each interaction as being either trustworthy or untrustworthy based on the degree of its satisfaction on the behavior of the agents involved in the interaction. Upon the receipt of a recommendation request from recommender system $r$ regarding agent $a_i$, agent $a_j$ has the choice to decide on whether to participate in the recommendation process or not. If $a_j$ accepts to participate, it will train the decision tree classifier on its dataset to predict the trustworthiness of $a_i$ based on the potential similarities between the specifications of $a_i$ and those of the agents that $a_j$ has previously dealt with (i.e., content-based recommendation). Based on the results of the machine learning classifier, agent $a_j$ recommends agent $a_i$ as being either trustworthy or untrustworthy. To avoid biased recommendations, the recommender system collects recommendations from multiple agents and aggregates them using DST to come up with a final aggregate decision that is resilient to dishonesty. Since the performance of DST is greatly dependent on the credibility of the parties giving the judgements, the third step involves updating the credibility scores of the participating agents on the basis of the convergence/divergence of their opinions w.r.t the final judgement given by DST. The proposed recommendation system is depicted in Algorithm \ref{Algo:Bootstrapping} (executed by the recommender system).
%Note that users could be incentivized to participate in the endorsement process in order to be able to make inquiries themselves when they encounter newly deployed items. In particular, inspired by an approach in \cite{wahab2016towards}, we could restrict each user to a limited number of inquiries it can make initially, where this number is increased whenever the user participates in an endorsement process, provided that it also maintains a good credibility score. Thus, over time, agents that refrain from reporting will get their inquiries drained and be unable to make further requests. In contexts such as cloud computing services, agents will have a desire to have reciprocation and trust values will be shared.

%Finally, to incentivize users to participate in the endorsement process, we restrict the number of endorsement inquiries that each user can initially from any other user to a certain limited number. This number then gets increased whenever the user participates in an endorsement process, provided that she also maintains a good credibility score. This way, over the time, the users that refrain from submitting endorsements will get their number of inquiries drained and hence won't be able to make further endorsement inquiries.
%To incentivize users to submit endorsements, all participating users are entered into a monthly draw for a chance to win a variety of prizes (e.g., prepaid credit cards, movie tickets), where this concept is widely adopted by many survey and department store companies (e.g., AskingCanadians, Walmart) to obtain users' opinions.

\begin{algorithm}
       \caption{Recommendation Algorithm}
\label{Algo:Bootstrapping}
        \begin{algorithmic}[1]
        \small
\makeatletter\setcounter{ALG@line}{0}\makeatother
\State \textbf{Input}: agent $a_i$ subject to recommendation
%\State \textbf{Input}: Request type $T\in \{\text{urgent}, \text{non-urgent}\}$
\State \textbf{Input}: Set $A$ of agents eligible to participate in the recommendation process
\State \textbf{Output}: Recommendation decision $R(r,a_i)$ of recommender system $r$ on agent $a_i$
\makeatletter\setcounter{ALG@line}{3}\makeatother
\Procedure{Recommendation}{}
%\State \textbf{repeat}
%\State \textbf{if} $T=\text{urgent}$ \textbf{then}
%\State \hspace{\algorithmicindent}Send bootstrapping request to currently online agents
%\State \textbf{else}
\State Broadcast the recommendation request to the set $A$ of agents
%\State \textbf{end if}
\State \textbf{for} each agent $a\in A$ \textbf{do}
%\State \hspace{\algorithmicindent}Verify the identity of $u$ using uport
\State \hspace{\algorithmicindent}Agent $a$ trains the decision tree classifier %using Eqs. \eqref{eq:Entropy}, \eqref{eq:WeightedEntropy}, and \eqref{eq:InformationGain}
\State \hspace{\algorithmicindent}Agent $a$ derives the recommendation on agent $a_i$ using decision tree
%\State \hspace{\algorithmicindent}\hspace{\algorithmicindent}\hspace{\algorithmicindent} if $N(s)$
\State \textbf{end for}
\State Use Eq. \eqref{eq:DSTrust} to compute the belief in $a_i$'s trustworthiness $\theta^{a_i}_r(T)$
\State Use Eq. \eqref{eq:DSMalicious} to compute belief in $a_i$'s untrustworthiness $\theta^{a_i}_r(N)$
%\State \hspace{\algorithmicindent} Use Eq. \eqref{eq:DSUncertain} to compute the belief $\theta^{i}(T)$ in item $i$ being either trustworthy or untrustworthy.
\State \textbf{if} $\theta^{a_i}_r(T) >  \theta^{a_i}_r(N)$ \textbf{then}
\State \hspace{\algorithmicindent} $R(r,a_i)$=trustworthy
\State \textbf{else}
\State \hspace{\algorithmicindent} $R(r,a_i)$=untrustworthy
\State \textbf{end if}
\State Update the credibility score of each agent $a\in A$ using Eq. \eqref{eq:CredibilityUpdateBoot}
\EndProcedure%
\end{algorithmic}
\end{algorithm}

\subsection{Recommendation Collection}
When an agent receives a recommendation request, it has the choice to either participate in the process or not. This voluntary aspect of participation is a building block in our solution to ensure fairness for both advisors and agents subject to recommendation. For example, some agents might not be willing to spend some time and resources helping other agents make choices. Moreover, some agents might not have sufficient accuracy (determined by the machine learning technique), lacking any similarity between the specifications of the agents dealt with and those of the agent being recommended. Therefore, refraining from participating would be the best choice for such agents instead of giving inaccurate recommendations (thanks to the credibility update mechanism proposed in Section \ref{CredibilityUpdate}). In case agents agree to participate, they first use the decision tree technique to predict the behavior of the underlying agents and derive the appropriate recommendations. Note that decision tree has been chosen for the considered problem due to its lightweight nature which requires no heavy computations nor long training time. This is important to (1) incentivize agents to participate in the recommendation collection process since they would not be required to use significant amounts of their resources (e.g., battery) to derive recommendations, and (2) minimize the time required to collect recommendations especially when it comes to urgent recommendation requests which require prompt answers.

\subsection{Aggregation to discourage dishonesty}
The purpose of this phase is to aggregate the different recommendations collected as per the previous phase in a non-collusive manner, i.e., in such a way that is resilient to agents that submit misleading recommendations to promote/demote some other agents. To do so, DST, which is known for its power in combining observations coming from multiple sources having different levels of credibilities, is employed. It is true that DST has been used in many proposals for trust establishment purposes \cite{yu2002evidential,wang2009new}; however, the main difference between our recommendations aggregation mechanism and the existing DST-based trust establishment techniques is that our solution requires no predefined thresholds to make a final decision on whether the agent should be trusted or not. Specifically, contrary to the existing approaches whose performance is greatly dependent on a certain threshold, we propose to compute both the belief in an agent's trustworthiness and untrustworthiness and comparing them to arrive at a final decision. Moreover, we propose in this work to weigh each witness (recommendation) based on the credibility score of its issuer, to reflect the dynamism of recommenders' honesty.

Formally, let $\Omega=\{T, N, U\}$ be a set composed of three hypotheses representing the possible recommendations on a certain agent, where $T$ means trustworthy, $N$ means untrustworthy, and $U$ means uncertainty between trust and distrust. The basic probability assignment (bpa) $m_b^{a_i}(H)$ of a particular hypothesis $H$ given by agent $b$ on agent $a_i$ is proportional to the credibility score of $b$. Specifically, if agent $b$ having a credibility score of $\lambda$ has recommended agent $a_i$ as being trustworthy, then the bpa's of the different hypotheses are computed as follows: $m_b^{i}(T) = \lambda$, $m_b^{a_i}(N) = 0$, and $m_b^{a_i}(U) = 1-\lambda$. Otherwise, if $b$ recommends $i$ as being untrustworthy, then the bpa's of the hypotheses are computed as follows: $m_b^{a_i}(T) = 0$, $m_b^{a_i}(N) = \lambda$, and $m_b^{a_i}(U) = 1-\lambda$. Having defined the bpa's, the final aggregate belief function regarding a certain hypothesis $H$ is computed through summing up all the bpa's coming from different recommenders upholding this hypothesis $H$. The belief function that recommender system $r$ computes regarding agent $a_i$'s trustworthiness after having consulted two recommenders $b$ and $b^\prime$ is given in Eq. \eqref{eq:DSTrust}.
\vspace{-0.07cm}
\begin{equation}
\label{eq:DSTrust}
\footnotesize
\theta_{r}^{a_i}(T)= m_b^{a_i}(T)\oplus m_{b^\prime}^{a_i}(T)=\frac{1}{K} [m_b^{a_i}(T)m_{b^\prime}^{a_i}(T)+m_b^{a_i}(T)m_{b^\prime}^{a_i}(U)+m_b^{a_i}(U)m_{b^\prime}^{a_i}(T)]
\end{equation}
\vspace{-0.07cm}
Similarly, the belief function computed by $r$ regarding agent $a_i$'s untrustworthiness after having consulted two recommenders $b$ and $b^\prime$ is given in Eq. \eqref{eq:DSMalicious}.
\vspace{-0.07cm}
\begin{equation}
\label{eq:DSMalicious}
\footnotesize
\theta_{r}^{a_i}(N)= m_b^{a_i}(N)\oplus m_{b^\prime}^{a_i}(N)=\frac{1}{K} [m_b^{a_i}(N)m_{b^\prime}^{a_i}(N)+m_b^{a_i}(N)m_{b^\prime}^{a_i}(U)+m_b^{a_i}(U)m_{b^\prime}^{a_i}(N)]
\end{equation}
\vspace{-0.07cm}
Finally, the belief function computed by $r$ regarding agent $a_i$ being either trustworthy or untrustworthy (i.e., uncertainty) after having consulted two recommenders $b$ and $b^\prime$ is given in Eq. \eqref{eq:DSUncertain}.
\vspace{-0.07cm}
\begin{equation}
\label{eq:DSUncertain}
\footnotesize
\theta_{r}^{a_i}(U)= m_b^{a_i}(U)\oplus m_{b^\prime}^{a_i}(U)=\frac{1}{K} [m_b^{a_i}(U)m_{b^\prime}^{a_i}(U)], \text{\normalsize where:}
\end{equation}
\vspace{-0.07cm}
\begin{equation}
\footnotesize
\label{eq:k}
\centering
\hspace{.6 cm} K=\sum_{h \cap h^\prime =\emptyset} m_b^{a_i}(h)m_{b^\prime}^{a_i}(h^\prime)
\end{equation}
Note that the values produced by the different belief functions are real-valued numbers between $0$ and $1$, i.e., $\theta_{r}^{a_i}(T)$, $\theta_{r}^{a_i}(N)$, $\theta_{r}^{a_i}(U) \in [0,1]$.
Finally, the decision of the recommender system regarding a certain agent $a_i$ is taken by computing the beliefs in $a_i$'s trustworthiness $\theta_{r}^{a_i}(T)$ and untrustworthiness $\theta_{r}^{a_i}(N)$ and comparing them, i,e., if $\theta_{r}^{a_i}(T)>\theta_{r}^{a_i}(N)$, $i$ is deemed trustworthy; otherwise, $a_i$ is considered as being untrustworthy.

\subsection{Credibility Update Mechanism}
\label{CredibilityUpdate}
The credibility scores of the participating agents need to be constantly updated in order to maintain the authenticity of the recommendation process. That is, honest recommenders should get their credibility values increased and dishonest recommenders should undergo a decrease in their credibility values. We propose in Eq. \eqref{eq:CredibilityUpdateBoot} a credibility update mechanism using which the recommender system $r$ updates its credibility belief $\phi(r \rightarrow a)$ toward every agent $a$ that has submitted a recommendation on another agent $a_i$ upon the request of $r$.

\begin{equation}
\small
  \phi(r \rightarrow a)=\begin{cases}
    \min{(1,\phi(r \rightarrow a)+X)}, & \text{if C \ref{CondPos}} \\
    |\phi(r \rightarrow a)-Y|, & \text{if C \ref{CondNeg}}\\
  \end{cases}
  \label{eq:CredibilityUpdateBoot}
\end{equation}
where $X=\max{(\theta_{r}^{a_i}(T),\theta_{r}^{a_i}(N))}$, $Y=\min{(\theta_{r}^{a_i}(T),\theta_{r}^{a_i}(N))}$, and C \ref{CondPos} and C \ref{CondNeg} are two conditions such that:
\newcounter{counter}
\setcounter{counter}{0}
\begin{condition}
\label{CondPos}
\footnotesize
 $R(a,a_i)\in \{T\} \hspace{.02cm}\& \hspace{.02cm}\theta_{r}^{a_i}(T)>\theta_{r}^{a_i}(N)$ or $R(a,a_i)\in \{N\} \hspace{.02cm}\& \hspace{.02cm}\theta_{r}^{a_i}(T)<\theta_{r}^{a_i}(M)$
\end{condition}

%\newcounter{foo}[2]
\begin{condition}
\label{CondNeg}
\footnotesize
$R(a,a_i)\in \{T\} \hspace{.02cm}\& \hspace{.02cm}\theta^{a_i}_{r}(T)<\theta^{a_i}_{r}(N)$ or $R(a,a_i)\in \{N\} \hspace{.02cm}\& \hspace{.02cm}\theta^{a_i}_{r}(T)>\theta^{a_i}_{r}(N)$
\end{condition}

The main idea of Eq. \eqref{eq:CredibilityUpdateBoot} is to update the credibility score of each recommender agent $a$ proportionally to the difference between her submitted recommendation $R(a,a_i)$ on agent $a_i$ and the final decision yielded by the DST-based aggregation mechanism. In this way, the agents whose recommendations converge to the final decision of the recommender system receive an increase in their credibility scores and those whose recommendations are far from the final decision undergo a decrease in their credibility scores. This process is of prime importance to guarantee the honesty of the recommendation process since the performance of the DST aggregation technique is highly dependent on the credibility scores of the recommenders.

\subsection{Incentive Mechanism and Participation Motivation}
In order to motivate the agents to participate in the recommendation process, we propose in this section an incentive mechanism which links the participation of the agents with the number of inquiries that they are allowed to make. Initially, all agents have an equal amount of inquiries that they are allowed to make from any other agent. This amount is then updated during the recommendation process as shown in Eq. \eqref{eq:Reward}. Specifically, every certain period of time, the number of inquiries that an agent $x$ is allowed to make from any other agent $s$ get increased in terms of the number of inquiries coming from $s$ that $x$ has answered, as well as the credibility score of $x$ believed by $s$, i.e.,

\begin{equation}
Inq(x\rightarrow s)=Inq(x\rightarrow s)+(\left\vert{E(x\rightarrow s)}\right\vert+\left \lceil{\left\vert{E(x\rightarrow s)}\right\vert\times Cr(s\rightarrow x)}\right \rceil+1)
\label{eq:Reward}
\end{equation}

In Eq. \eqref{eq:Reward}, $Inq(x\rightarrow s)$ denotes the total number of inquiry requests that $x$ is allowed to make from $s$, $\left\vert{E(x\rightarrow s)}\right\vert$ denotes the number of recommendations that $x$ has answered in favor of $s$, and $Cr(s\rightarrow x)$ denotes the credibility score of agent $x$ believed by agent $s$. In this way, the agents that refuse to participate in the recommendation process would, over time, end up being unable to make any request from any other agent. In addition, by linking the number of inquiries with the credibility score of the recommender agent, we aim at motivating those agents to provide honest opinions.

\section{Experimental Evaluation}
\label{Sec:Simulations}
\subsection{Experimental Setup and Datasets}
We compare our solution with the MET recommendation-based trust model proposed in \cite{jiang2013evolutionary}. The objective of MET is to derive the optimal trust network that provides the most accurate estimation of sellers' reputation scores in duopoly environments. To carry out these experiments, we consider a similar environment to that considered in \cite{jiang2013evolutionary} by simulating three types of attacks that can be launched
by consulted agents (i.e, advisors) to mislead the recommendation-based trust establishment process (i.e., Sybil, camouflage, and whitewashing) and setting the percentage of attackers to $30\%$ of the total number of advisors. We compare both approaches in terms of Mean Absolute Error (MAE), which is computed as follows: $MAE(i)= \frac{\mid Trust(i)-\hat{Trust(i)}\mid}{\mid A\mid}$, where $Trust(i)$ is the actual trust score of item $i$, $\hat{Trust(i)}$ is the trust value of item $i$ estimated by the trust model, and $\mid A\mid$ is the number of consulted agents. In Sybil attacks, dishonest agents generate several fake identities in an attempt to manipulate the trust aggregation process by submitting a large number of dishonest recommendations. In camouflage attacks, attackers try to fool the trust system by providing honest recommendations in the beginning to build good credibility scores and then start to submit dishonest recommendations. In whitewashing, dishonest agents try to clear their bad credibility history through continuously creating new identities.

To conduct the experiments, real-world trust data from the Epinions\footnote{http://www.epinions.com/} large Web community are employed. Epinions allows users to express their opinions regarding a wide variety of items (e.g., movies, cars, etc.) in the form of numeric ratings from the interval $[1,5]$, with $1$ being the rating which represents the least satisfaction level and $5$ being the rating which represents a full satisfaction level. Epinions allows users as well to rate each other on the basis of the meaningfulness of their submitted ratings. The dataset consists of approximately $140,000$ items rated by $50,000$ users, where a total of $\approx 660,000$ reviews are collected \cite{massa2006trust}. We chose to use Epinions when comparing with other approaches thanks to its large-scale nature, which allows us to better test the generalizability of the studied solutions. The machine learning classifier has been trained on the dataset following the $k$-fold cross-validation approach (with $k=10$) \cite{refaeilzadeh2009cross}. Note finally that the experiments have been conducted using Matlab in a $64$-bit Windows $7$ environment on a machine equipped with an Intel Core i$7$-$4790$ CPU $3.60$ GHz Processor and $16$ GB RAM.

\subsection{Results and Discussion}

\begin{table}[!ht]
\begin{center}
\caption{\small Mean Absolute Error (MAE) comparison of items' initial trust estimation between our solution and MET \cite{jiang2013evolutionary}}
\label{table:MAEComparison}
\scalebox{0.90}{
\begin{tabular}{|c|c|c|}
 \hline\hline
&Our Solution&MET\\
\hline\hline
Sybil&$0.05\pm 0.02$&$0.09\pm 0.06$\\
\hline
Camouflage&$0.12\pm 0.01$& $0.01\pm 0.00$\\
\hline
Whitewashing&$0.06\pm 0.03$& $0.05\pm 0.03$\\
\hline\hline
\end{tabular}
}
\end{center}
\end{table}

We notice from Table \ref{table:MAEComparison} that our solution decreases the MAE compared to MET in the presence of Sybil attacks. In fact, MET is based on the idea of generating various networks of advisors with different trust values using evolutionary operators and then keeping the best network that consists of advisors having the highest trust values. However, even in the optimal network of advisors, there is still some chances of encountering a minority of advisors that might provide inaccurate recommendations. Worse, in the case of Sybil attacks, such a minority might even become a majority by creating a large number of fake identities. In such cases, MET provides no countermeasures against such advisors. On the other hand, our solution operates not only at the collection level but also at the recommendation aggregation level, thus providing an additional countermeasure against dishonesty. Specifically, in our approach, we broadcast the recommendation requests to the set of available advisors and allow these advisors to self-assess their degrees of accuracy prior to submitting their recommendations. Thereafter, we aggregate the recommendations coming from advisors having different levels of credibility using the Dempster-Shafer method, which is mainly influenced by the credibility of the agents, rather than their number. This makes our solution quite resilient to Sybil attacks and efficient even in scenarios in which dishonest advisors might form the majority \cite{wahab2016towards}.

On the other hand, in camouflage attacks, MET entails lower MAE compared to our solution. The reason is that in such a type of attacks, dishonest advisors initially provide honest recommendations to build up good credibility scores, prior to starting their dishonest recommendations. This makes our recommendation-based trust aggregation method, which is mainly influenced by the credibility scores of the advisors, to be vulnerable to such attackers for a short period of time (i.e., the period at which dishonest advisors switch their behavior and start providing misleading recommendations). When it comes to whitewashing attacks, our solution and MET show relatively similar resilience to such attacks. In fact, MET keeps only the network of advisors with the most suitable trust values according to some evolutionary operators, which makes it hard for whitewashing agents to get into these networks. In our solution, even though some dishonest agents might clear their bad credibility history and rejoin the network again, such newcomer agents aren't expected to have high credibility values as compared to those honest agents who have strived to build and retain high credibility scores. Consequently, the presence of camouflage attackers does not have a significant impact on the performance of our solution.

\begin{figure*}
        \centering
        \begin{subfigure}{0.31\textwidth}
                \scalebox{0.50}{\includegraphics[bb=220 514 418 725]{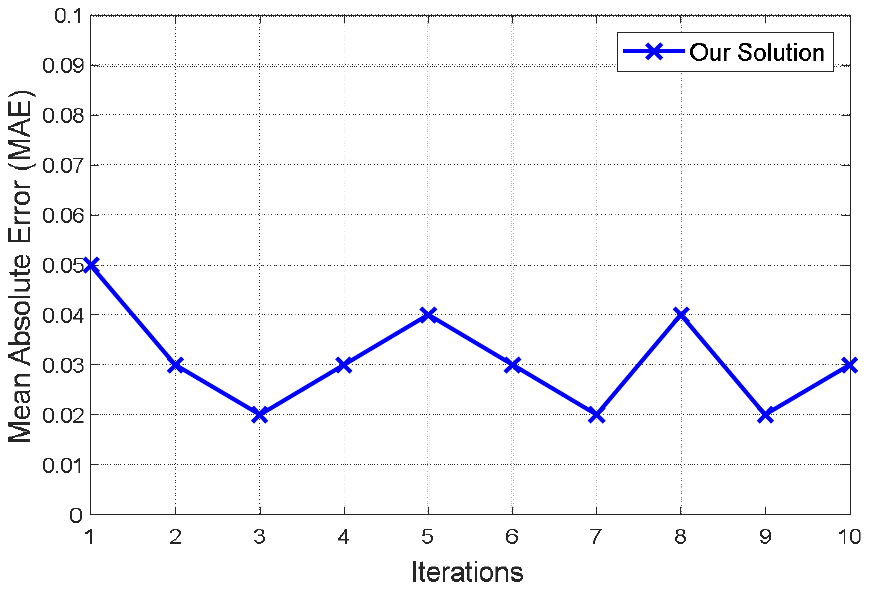}}
                \vspace{-0.2cm}
                %\newline
                \caption{Sybil}
                \label{Fig:SybilEffect}
        \end{subfigure}%
        %\quad
        \begin{subfigure}{0.31\textwidth}
                \scalebox{0.50}{\includegraphics[bb=190 514 418 735]{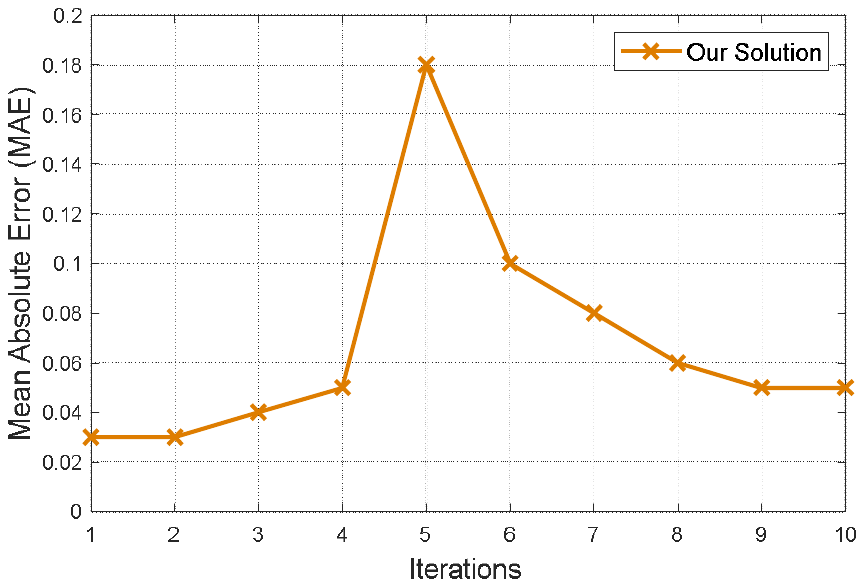}}
                \vspace{-0.6cm}
                \caption{Camouflage}
                \label{Fig:CamouflageEffect}
        \end{subfigure}%
        %\quad%\qquad%\qquad
        \begin{subfigure}{0.31\textwidth}
                \scalebox{0.48}{\includegraphics[bb= 155 514 418 730]{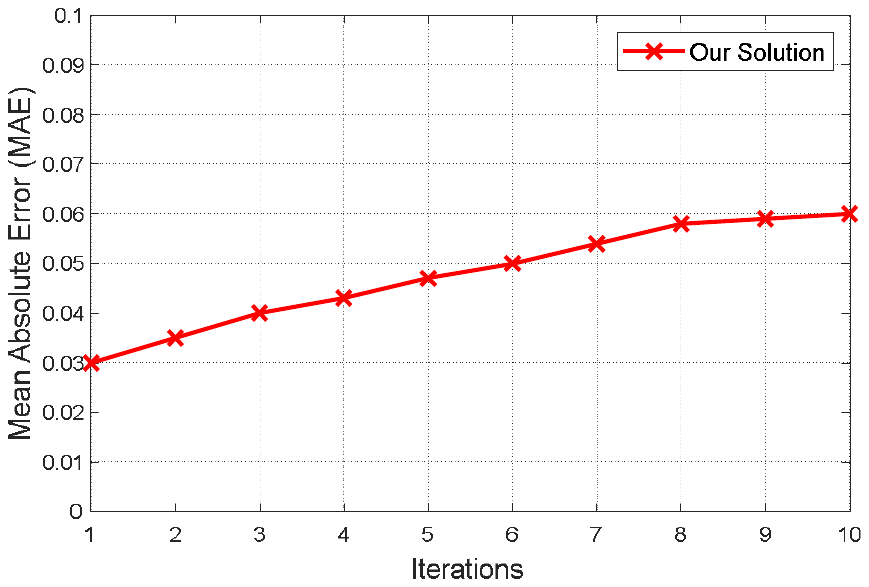}}
                \vspace{-0.6cm}
                \caption{Whitewashing}
                \label{Fig:WhitewashingEffect}
        \end{subfigure}
        \caption{The effects of Sybil, camouflage, and whitewashing attacks on the MAE entailed by our solution}\label{Fig:Attacks}
\end{figure*}

In Fig. \ref{Fig:Attacks}, we study in more detail the effects of each of the three types of attacks on our solution. We can notice from Fig. \ref{Fig:SybilEffect} that Sybil attacks have no effect on our solution, for the reasons mentioned above. From Fig. \ref{Fig:CamouflageEffect}, we can see that up to the fifth iteration (the period during which camouflage attackers provide honest recommendations to gain high credibility scores), the MAE entailed by our solution is low. At the fifth iteration (the time moment at which camouflage attackers start to change their behavior by providing dishonest recommendations), the MAE of our solution is reported to be relatively high (i.e., $0.18$). Starting from the sixth iteration, our solution starts to recognize the camouflage attackers and decrease their credibility, thus leading to gradually improving the performance and decreasing the MAE. Finally, from \ref{Fig:WhitewashingEffect}, we notice that whitewashing attacks have a small effect on the performance of our solution for the reasons mentioned in the previous paragraph.

\section{Conclusion and Future Work}
\label{sec:conclusion}
Ensuring the honesty of recommendations is a building block for the success of any recommendation-based trust model. We proposed in this paper a mechanism to deal with dishonest recommendations at both the collection and processing levels. Experimental results on a real-life dataset revealed that our solution considerably increases the accuracy of recommendations in the presence of Sybil attackers. Our solution shows also significant resilience to whitewashing and camouflage attacks.

Three issues are important to continue to consider for future work. The first is that of identity management, which is important to verify the identities of the recommenders and prevent identity impersonation and/or duplication. To address this issue, we plan to integrate blockchain-based solutions such as uport\footnote{https://www.uport.me/} into our solution. Another issue is addressing cases where peers have subjective differences. In the future, we plan to extend our solution to support situations wherein users might use different evaluation functions, thus learning these functions in a manner similar to that of the BLADE system \cite{regan2006bayesian}. A final concern is to continue to incentivize peers to provide recommendations. This is an ongoing concern for recommender systems, especially in crowdsourced environments.

\bibliographystyle{splncs04}
\bibliography{sample-bibliography}
\nocite{*}

\end{document}